\documentclass[runningheads]{llncs}
\usepackage{graphicx}
\usepackage[T1]{fontenc}
\usepackage{caption}
\usepackage{subcaption}

\usepackage{booktabs}
\newcommand{\ra}[1]{\renewcommand{\arraystretch}{#1}}

\begin{document}
\title{diveXplore 6.0: ITEC's Interactive Video Exploration System at VBS 2022}
\titlerunning{diveXplore 6.0 at VBS 2022}
%

\author{Andreas Leibetseder \and 
 Klaus Schoeffmann} 

\authorrunning{A. Leibetseder et al.}
%
\institute{Klagenfurt University, Institute of Information Technology (ITEC), \\ Klagenfurt, Austria
\email{\{aleibets,ks\}@itec.aau.at}}
\maketitle              
\begin{abstract}


Continuously participating since the sixth Video Browser Showdown (VBS2017), diveXplore is a veteran interactive search system that throughout its lifetime has offered and evaluated numerous features. After undergoing major refactoring for the most recent VBS2021, however, the system since version 5.0 is less feature rich, yet, more modern, leaner and faster than the original system. This proved to be a sensible decision as the new system showed increasing performance in VBS2021 when compared to the most recent former competitions. With version 6.0 we reconsider shot segmentation, map search and introduce new features for improving concept as well as temporal context search.

\end{abstract}


\section{Introduction}

The recurring team competition Video Browser Showdown~\cite{VBS2014,Lokoc2018,lokoc2019interactive,rossetto2020interactive} (VBS) aims at improving the quick interactive search of large video databases, in particular V3C1 and V3C2~\cite{rossetto2021insights,berns2019v3c1}. Several competing teams develop custom retrieval systems using a known dataset in order to solve several types of tasks within a short time frame at an international live event: visual and textual Known Item Search (KIS) for retrieving specific unique items and Ad-hoc Video Search (AVS) for finding multiple items with a certain characteristic. Moreover, the developed systems typically are used by persons familiar with them (experts) as well as inexperienced users (novices). The diveXplore system~\cite{schoeffmann2017collaborative,primus2018itec,leibetseder2018sketch,schoeffmann2019autopiloting,diveXplore2020,leibetseder2021less} annually participates in these competitions since VBS2017, ever growing in complexity. With version 5.0~\cite{leibetseder2021less} we, therefore, decided to refactor the system optimizing its responsiveness and speed, while reducing its features, achieving the 12$^{th}$ rank out of 18 teams in VBS2021. In version 6.0 of the system, we aim at improving its underlying shot segmentation as well as its capabilities of temporal search. In the following we detail the system and describe its improvements for VBS2022.

\section{diveXplore 6.0}



\subsection{Architecture}
\label{sec:architecture}

\begin{figure*}[htb!]
  \centering
  \includegraphics[width=\linewidth]{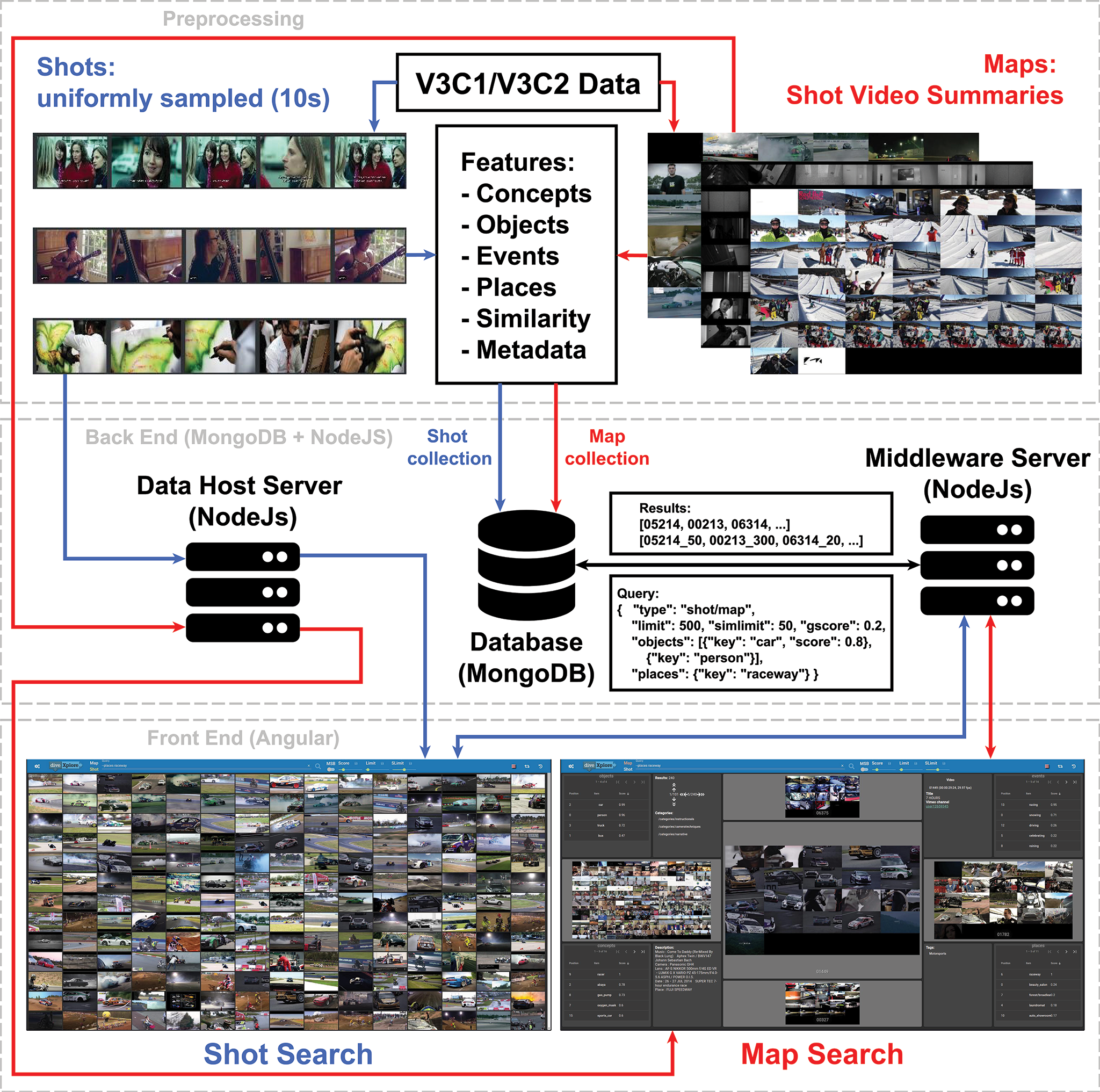}
  \caption{diveXplore architecture}
  \label{fig:architecture}
\end{figure*}

The interactive retrieval system diveXplore operates on metadata and preprocessed features extracted from video keyframes, i.e. representative frames belonging to custom-segmented consecutive video scenes. It is built in a modular fashion and based on web-technologies enabling a user to issue combinable queries in two different modes: shot and map search. Figure~\ref{fig:architecture} illustrates the system's main components grouped by type: preprocessing, back and front end. 

Generally, both of the diveXplore's search modes rely on its custom shot segmentation of uniformly sampling the source datasets into ten-second scenes. However, they differ in their targeted amount of data: shot search is concerned with features belonging to individual shot keyframes, while map search operates on a video basis: results are matched by feature frequency in all shots of particular videos. Additionally, this mode includes similarity search among video summary images (maps), which is determined via calculating their feature vector similarities. Table~\ref{tab:features} lists and briefly describes all of our utilized deep features.

\begin{table}[tbp!]
    \centering
    \ra{1.3}
    \caption{diveXplore features}
    \label{tab:features}
    \begin{tabular}{@{}llcl@{}}\toprule
        \multicolumn{1}{l}{\textbf{Type}} &%
        \multicolumn{1}{l}{\textbf{Model}} &%
        \multicolumn{1}{c}{\textbf{Ref.}} &%
        \multicolumn{1}{l}{\textbf{Description}} \\
        \midrule
        Concepts & Inception v3 & \cite{DBLP:conf/cvpr/SzegedyVISW16} & 1000 ImageNet categories. \\
        Objects & YOLO v4 & \cite{bochkovskiy2020yolov4} & 80 MS COCO categories. \\
        Events & Moments in Time & \cite{monfortmoments} & 304 Moments in Time action events. \\
        Places & Places365 & \cite{zhou2017places} & 365 individual places. \\
        Similarity (map) & Inception v3 & \cite{DBLP:conf/cvpr/SzegedyVISW16} & Based on last fully connected (FC) layer. \\
        \bottomrule
    \end{tabular}
\end{table}


Most extracted deep features, i.e. concepts, objects and places, are extracted from shot keyframes, yet, action events require temporal context and are therefore determined using the full ten-second segments. In addition to hosting shots, maps and videos on a Node.js (\url{https://nodejs.org}) data host server, we store shot as well as map feature collections on a MongoDB (\url{https://mongodb.com}) database, which is accessible using a middleware Node.js query server. As portrayed in Figure~\ref{fig:architecture}, queries as well as responses are formulated via POST request containing a JSON body, which allows for complex query combinations in both modes based on features and confidence thresholds. In this manner, the search interface can be deployed rather easily as it is possible to host and share all of the underlying data through dedicated, potentially more powerful servers.







\subsection{Interface}
\label{sec:interface}



\begin{figure*}[htbp!]
    \centering
    \begin{subfigure}{0.48\textwidth}
      \centering
      \includegraphics[width=\linewidth]{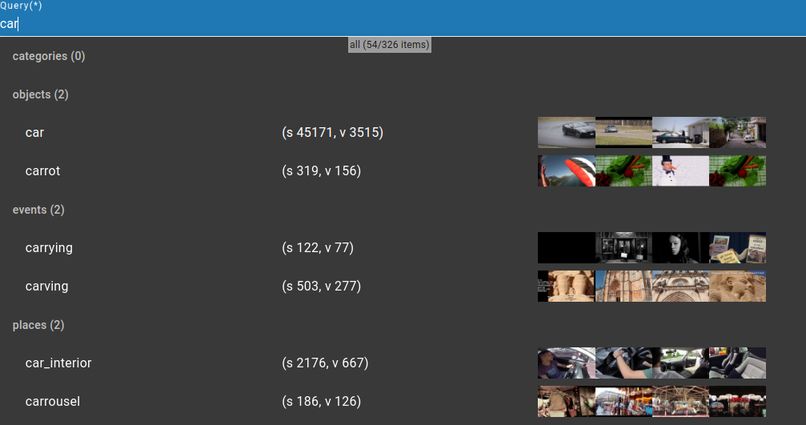}
      \caption{Top autocomplete suggestions for 'car'}
      \label{fig:autocomplete}
    \end{subfigure}%
    \hspace{0.5em}
    \begin{subfigure}{0.48\textwidth}
      \centering
      \includegraphics[width=\linewidth]{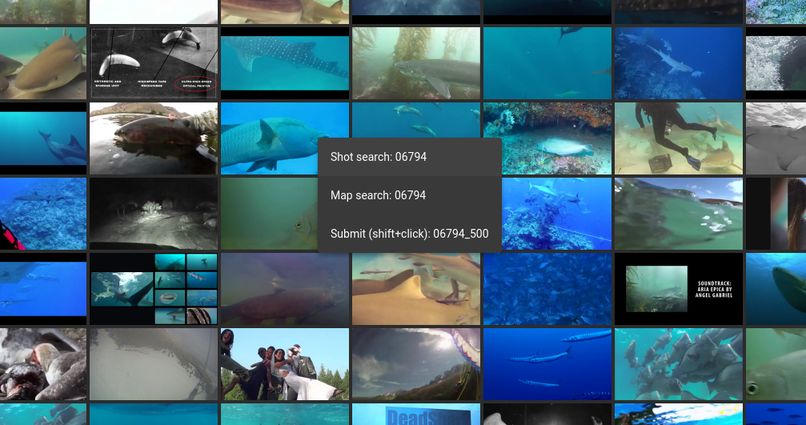}
      \caption{Shot context menu}
      \label{fig:context_menu}
    \end{subfigure}

    \begin{subfigure}{0.48\textwidth}
      \centering
      \includegraphics[width=\linewidth]{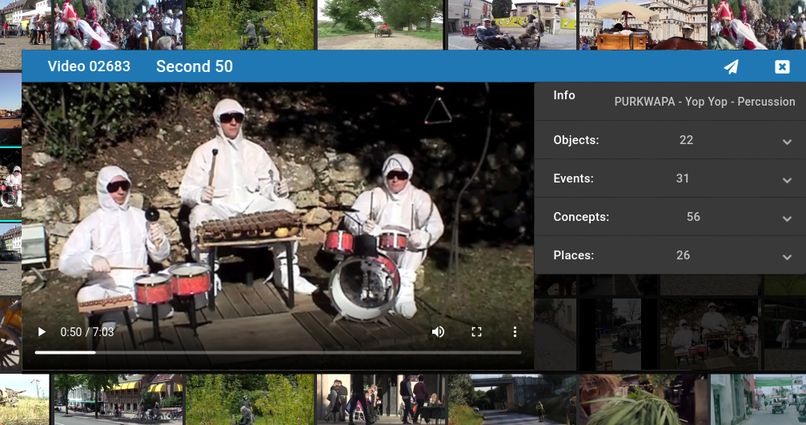}
      \caption{Video overlay with shot details}
      \label{fig:video_overlay}
    \end{subfigure}
    \hspace{0.5em}    
    \begin{subfigure}{0.48\textwidth}
      \centering
      \includegraphics[width=\linewidth]{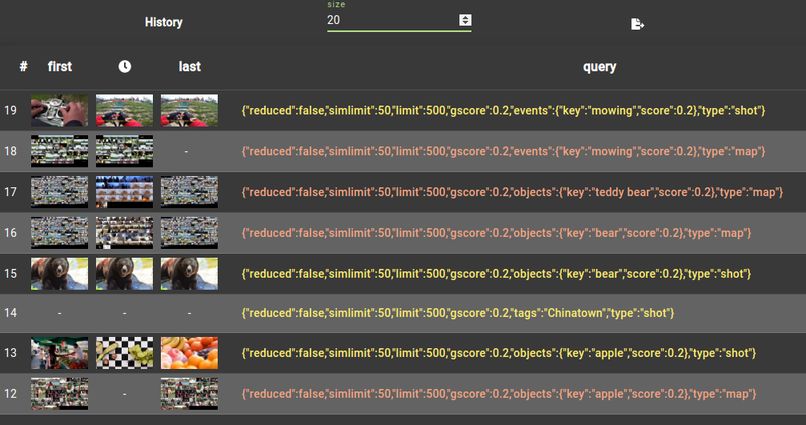}
      \caption{History with browsed shots}
      \label{fig:history}
    \end{subfigure}%
    
  \caption{diveXplore interface components}
  \label{fig:features_a}
\end{figure*}

As depicted the front end part of Figure~\ref{fig:architecture}, diveXplore includes two different interfaces accommodating shot and map search mode. Despite displaying results quite differently, both share the same search bar (top blue area), which enables entering a series of combinable search terms. E.g., for creating a database request similar to the one shown in Figure~\ref{fig:architecture}, a user enters the following list of search terms in a bash-like manner: '\texttt{--objects car (0.8), person --places raceway}'. Since entering many search terms this way could be tedious, the system offers smart autocompletion functionalities: besides listing available concepts and their shortcuts, as a user starts typing the autocompletion bar is updated to contain all relevant terms together with their concept categories. It even includes the frequency and example images of the suggested search concepts, as is in Figure~\ref{fig:autocomplete} demonstrating autocomplete with the example input '\texttt{car}'.

For result display (cf. Figure~\ref{fig:architecture}), shot search offers a simple, traditional view: it displays a grid of shots ordered by most to least relevant. This mode is useful for quickly scanning a large variety of shots regardless of the videos they originate from, which, for instance, is convenient for AVS tasks. The map view, however, offers a novel way of result browsing. By navigating aforementioned video summary maps, a user is able to quickly overview a video's complete content. Only one map is inspected at any given moment and all the information in the corner areas always correspond to that map, which is the largest one centered in the middle of the screen. The other maps are previews of the previous and next maps for two exploration modes: horizontal navigation traverses the list of result videos from highest to lowest rank according to feature frequency and vertical navigation enables a user to view similar videos to currently selected horizontal map. We've made it as easy as possible to switch from map to shot search -- a single click suffices for issuing the current query using the respective other mode in a new browser tab. Also, as shown in Figure~\ref{fig:context_menu}, the system implements convenience context menus to search for a shot in each available mode. Furthermore, the system provides the user with a video preview feature that can be activated by selecting any shot. As demonstrated in Figure~\ref{fig:video_overlay}, this overlay not only includes general information such as video description and tags but also its topmost objects, concepts, events and places. This video browser enables submitting much more precise timestamps for KIS tasks compared to simply submitting shot keyframes. Finally, diveXplore offers an extensive history feature (cf. Figure~\ref{fig:history}) which not only precisely records the issued queries color-coded by their type, it also stores shot thumbnails according to the user's browsing activity, i.e. first, last and longest inspected shot.





\subsection{Improvements}

\begin{table}[tbp!]
    \centering
    \ra{1.3}
    \caption{System improvements impacting front end (fe) and back end (be)}
    \label{tab:improvements}
    \begin{tabular}{@{}lll@{}}\toprule
        \multicolumn{1}{l}{\textbf{Improvement}} &%
        \multicolumn{1}{l}{\textbf{Impact}} &%
        \multicolumn{1}{l}{\textbf{Description}} \\
        \midrule
        1s segmentation & be, fe & Improving KIS performance. \\
        alternative map search & be & Improving map retrieval accuracy. \\
        OCR, STT & be & New search features (e.g.~\cite{kay2007tesseract},~\cite{povey2011kaldi}). \\
        temporal search & be, fe & Improving result precision. \\
        \bottomrule
    \end{tabular}
\end{table}


Our system improvements are listed in Table~\ref{tab:improvements}. Although not having a big impact on AVS search, the rather long 10s shot segmentation seems to greatly impede KIS tasks -- seemingly simple tasks could not be found, even after the assigned search time. This especially happens in case the targeted video segments are very short. Therefore, we reduce the keyframe sampling interval to one-second for VBS2022. Furthermore, since merely using feature frequency for map retrieval seems to have detrimental effects on some queries, we explore alternative ways to match those features to improve result accuracy. Additionally, we introduce useful and proven successful features such as Optical Character Recognition (OCR) and Speech-to-Text (STT). Lastly, while temporal events are useful to retrieve activities, the system requires more sophisticated means for temporal context search, which also will include single keyframe-based features. 

\section{Conclusion}

We introduced diveXplore 6.0 as a competing system at VBS2022. While some features proved to be helpful since refactoring the system, we continue this process by reducing the shot sampling interval, improving map search, adding proven useful features and offering sophisticated temporal concept search.

\section*{Acknowledgments}

{\small
    This work was funded by the FWF Austrian Science Fund under grant P 32010-N38.
}


%
%
%
\bibliographystyle{splncs04}
{
\scriptsize
\bibliography{refs.bib}

\begin{thebibliography}{10}
\providecommand{\url}[1]{\texttt{#1}}
\providecommand{\urlprefix}{URL }
\providecommand{\doi}[1]{https://doi.org/#1}

\bibitem{berns2019v3c1}
Berns, F., Rossetto, L., Schoeffmann, K., Beecks, C., Awad, G.: V3c1 dataset:
  An evaluation of content characteristics. In: Proc. of the 2019 on Intl.
  Conf. on Multimedia Retrieval. pp. 334--338. ACM (2019)

\bibitem{bochkovskiy2020yolov4}
Bochkovskiy, A., Wang, C.Y., Liao, H.Y.M.: Yolov4: Optimal speed and accuracy
  of object detection. CoRR  \textbf{abs/2004.10934} (2020),
  \url{https://arxiv.org/abs/2004.10934}

\bibitem{kay2007tesseract}
Kay, A.: Tesseract: an open-source optical character recognition engine. Linux
  Journal  \textbf{2007}(159), ~2 (2007)

\bibitem{leibetseder2018sketch}
Leibetseder, A., Kletz, S., Schoeffmann, K.: Sketch-based similarity search for
  collaborative feature maps. In: Intl. Conf. on Multimedia Modeling. pp.
  425--430. Springer (2018)

\bibitem{diveXplore2020}
Leibetseder, A., M{\"u}nzer, B., Primus, J., Kletz, S., Schoeffmann, K.:
  divexplore 4.0: The itec deep interactive video exploration system at
  vbs2020. In: Ro, Y.M., Cheng, W.H., Kim, J., Chu, W.T., Cui, P., Choi, J.W.,
  Hu, M.C., De~Neve, W. (eds.) MultiMedia Modeling. pp. 753--759. Springer
  Intl. Publishing, Cham (2020)

\bibitem{leibetseder2021less}
Leibetseder, A., Schoeffmann, K.: Less is more-divexplore 5.0 at vbs 2021. In:
  Intl. Conf. on Multimedia Modeling. pp. 455--460. Springer (2021)

\bibitem{Lokoc2018}
Lokoc, J., Bailer, W., Schoeffmann, K., Muenzer, B., Awad, G.: On influential
  trends in interactive video retrieval: Video browser showdown 2015-2017. IEEE
  Transactions on Multimedia pp.~1--1 (2018). \doi{10.1109/TMM.2018.2830110}

\bibitem{lokoc2019interactive}
Loko\v{c}, J., Koval\v{c}\'{\i}k, G., M\"{u}nzer, B., Sch\"{o}ffmann, K.,
  Bailer, W., Gasser, R., Vrochidis, S., Nguyen, P.A., Rujikietgumjorn, S.,
  Barthel, K.U.: Interactive search or sequential browsing? a detailed analysis
  of the video browser showdown 2018. ACM Trans. Multimedia Comput. Commun.
  Appl.  \textbf{15}(1),  29:1--29:18 (Feb 2019). \doi{10.1145/3295663},
  \url{http://doi.acm.org/10.1145/3295663}

\bibitem{monfortmoments}
Monfort, M., Vondrick, C., Oliva, A., Andonian, A., Zhou, B., Ramakrishnan, K.,
  Bargal, S.A., Yan, T., Brown, L.M., Fan, Q., Gutfreund, D.: Moments in time
  dataset: One million videos for event understanding. {IEEE} Transactions on
  Pattern Analysis and Machine Intelligence  \textbf{42}(2),  502--508 (2020),
  \url{https://doi.org/10.1109/TPAMI.2019.2901464}

\bibitem{povey2011kaldi}
Povey, D., Ghoshal, A., Boulianne, G., Burget, L., Glembek, O., Goel, N.,
  Hannemann, M., Motlicek, P., Qian, Y., Schwarz, P., et~al.: The kaldi speech
  recognition toolkit. In: IEEE 2011 workshop on automatic speech recognition
  and understanding. No.~CONF, IEEE Signal Processing Society (2011)

\bibitem{primus2018itec}
Primus, M.J., M{\"u}nzer, B., Leibetseder, A., Schoeffmann, K.: The itec
  collaborative video search system at the video browser showdown 2018. In:
  Intl. Conf. on Multimedia Modeling. pp. 438--443. Springer (2018)

\bibitem{rossetto2020interactive}
{Rossetto}, L., {Gasser}, R., {Lokoč}, J., {Bailer}, W., {Schoeffmann}, K.,
  {Muenzer}, B., {Souček}, T., {Nguyen}, P.A., {Bolettieri}, P.,
  {Leibetseder}, A., {Vrochidis}, S.: Interactive video retrieval in the age of
  deep learning – detailed evaluation of vbs 2019. IEEE Transactions on
  Multimedia  \textbf{23},  243--256 (2021). \doi{10.1109/TMM.2020.2980944}

\bibitem{rossetto2021insights}
Rossetto, L., Schoeffmann, K., Bernstein, A.: Insights on the v3c2 dataset.
  arXiv preprint arXiv:2105.01475  (2021)

\bibitem{VBS2014}
Schoeffmann, K.: A user-centric media retrieval competition: The video browser
  showdown 2012-2014. MultiMedia, IEEE  \textbf{21}(4),  8--13 (Oct 2014).
  \doi{10.1109/MMUL.2014.56}

\bibitem{schoeffmann2019autopiloting}
Schoeffmann, K., M{\"u}nzer, B., Leibetseder, A., Primus, J., Kletz, S.:
  Autopiloting feature maps: The deep interactive video exploration
  (divexplore) system at vbs2019. In: Intl. Conf. on Multimedia Modeling. pp.
  585--590. Springer (2019)

\bibitem{schoeffmann2017collaborative}
Schoeffmann, K., Primus, M.J., Muenzer, B., Petscharnig, S., Karisch, C., Xu,
  Q., Huerst, W.: Collaborative feature maps for interactive video search. In:
  Intl. Conf. on Multimedia Modeling. pp. 457--462. Springer (2017)

\bibitem{DBLP:conf/cvpr/SzegedyVISW16}
Szegedy, C., Vanhoucke, V., Ioffe, S., Shlens, J., Wojna, Z.: Rethinking the
  inception architecture for computer vision. In: Conf. on Computer Vision and
  Pattern Recognition. pp. 2818--2826. {IEEE} (2016),
  \url{https://doi.org/10.1109/CVPR.2016.308}

\bibitem{zhou2017places}
Zhou, B., Lapedriza, A., Khosla, A., Oliva, A., Torralba, A.: Places: A 10
  million image database for scene recognition. {IEEE} Transactions on Pattern
  Analysis and Machine Intelligence  \textbf{40}(6),  1452--1464 (2018),
  \url{https://doi.org/10.1109/TPAMI.2017.2723009}

\end{thebibliography}
}

\end{document}